\documentclass[aps, prl, showpacs, twocolumn]{revtex4}
\usepackage{graphicx}
\usepackage{dcolumn}

\begin{document}
\title{Effects of Two Energy Scales in Weakly Dimerized Antiferromagnetic Quantum Spin Chains}

\author{A. Br\"uhl$^1$, B. Wolf$^1$, V. Pashchenko$^1$, M. Anton$^1$, C. Gross$^1$,
 W. Assmus$^1$, R. Valenti$^2$, S. Glocke$^3$, A. Kl\"umper$^3$,T. Saha-Dasgupta$^4$, B. Rahaman$^4$, and M. Lang$^1$}

\address{$^1$Physikalisches Institut, Universit\"{a}t Frankfurt,
D-60438 Frankfurt, Germany}
\address{$^2$Institut f\"{u}r Theoretische Physik, Universit\"{a}t Frankfurt,
D-60438 Frankfurt(M), Germany}
\address{$^3$Theoretische Physik, Universit\"{a}t Wuppertal, 42097 Wuppertal, Germany }
\address{$^4$S.N. Bose National Centre for Basic Sciences, Salt Lake City, Kolkata 700098, India}

\date{\today}

\begin{abstract}
By means of thermal expansion and specific heat measurements on
the high-pressure phase of (VO)$_2$P$_2$O$_7$, the effects of two
energy scales of the weakly dimerized antiferromagnetic $S$ = 1/2
Heisenberg chain are explored. The low energy scale, given by the
spin gap $\Delta$, is found to manifest itself in a pronounced
thermal expansion anomaly. A quantitative analysis, employing
T-DMRG calculations, shows that this feature originates from
changes in the magnetic entropy with respect to $\Delta$,
$\partial S^{m}/
\partial \Delta$. This term, inaccessible by specific
heat, is visible only in the weak-dimerization limit where it
reflects peculiarities of the excitation spectrum and its
sensitivity to variations in $\Delta$.

\end{abstract}


\maketitle

In recent years, many interesting and unexpected effects have been
discovered in low-dimensional antiferromagnetic (afm) quantum-spin
systems. Prime examples include the Haldane gap for spin $S$ = 1
chains, magnetization plateaux in coupled-dimer systems, and the
different ground states for even- and odd-leg ladder systems, see,
e.g.\,\cite{Haldane83, Affleck89, Dagotto96, Oshikawa97}. These
phenomena reflect the intricate many-body quantum character of the
systems which renders advanced mathematical methods necessary to
gain a quantitative understanding of the experimental
observations. In this communication, we want to address another
non-trivial collective phenomenon, overlooked in the past, of the
effects of two different energy scales caused by a weak
alternation in the spin-spin interaction of an $S$ = 1/2 afm
Heisenberg chain. These scales determine the physical properties
of the interacting system at different temperatures. If $J_1$ and
$J_2$ ($< J_{1}$) denote the alternating afm coupling constants
between nearest neighboring spins along the chain, we define $J_1$
as the large scale and the spin gap $\Delta$, induced by any
amount of alternation \cite{Bonner79} $\tilde{\alpha} < 1$ with
$\tilde{\alpha} \equiv J_2/J_1$, as the small one. The dominant
feature, governing the magnetic and thermodynamic properties at
elevated temperatures, is connected with $J_1$. In the magnetic
susceptibility $\chi$, for example, the maximum due to short-range
order is located at $k_B T_{\chi}^{max} = 0.64 J_1$, practically
independent of $\tilde{\alpha}$ \cite{Johnston00}. However as will
be shown below, the small energy scale introduces a second
distinct anomaly \cite{Signatures}, which can be extraordinarily
large in the coefficient of thermal expansion. The feature
observed here is no longer proportional to that in the specific
heat, viz., the so-called Gr\"{u}neisen-scaling \cite{Barron80}
does not apply.

For the study of the effects of two energy scales in the
interesting weak-dimerization regime, the high-pressure variant of
(VO)$_2$P$_2$O$_7$ (HP-VOPO) appears particularly well suited.
HP-VOPO, first synthesized by Azuma \textit{et
al.}\,\cite{Azuma99}, comprises one kind of $S$ = 1/2 (V$^{4+}$)
chains with alternating afm Heisenberg interactions \cite{Azuma99,
Johnston01}, as opposed to the ambient-pressure (AP) variant,
where two slightly different chain types were identified
\cite{Garrett97a, Johnston01}. No magnetic ordering has been found
in HP-VOPO down to 1.5\,K $\sim 0.01 J_{1}/k_{B}$, the lowest
temperature studied so far, indicating that interchain
interactions are weak.

To verify that the model of an alternating afm Heisenberg chain is
appropriate for HP-VOPO, we carried out {\it ab initio} density
functional (DFT) calculations in the generalized gradient
approximation using an LMTO basis. Our analysis of the DFT results
in terms of the Nth-order muffin-tin orbital-based downfolding
technique \cite{Dasgupta00}, shows that the dominant V-V effective
hopping integrals are the intrachain hopping via the PO$_4$ unit,
$t_1$ = 0.114\,eV, followed by the nearest neighbor V-V intrachain
hopping $t_2$ = 0.10\,eV. The next non-negligible V-V hopping is
along the $b$-axis, $t_b$ = 0.02\,eV. Since all other possible V-V
interaction paths are more than one order of magnitude smaller
than $t_{1,2}$, this analysis clearly confirms the
alternating-chain nature of HP-VOPO.

Single crystals of HP-VOPO were prepared by transforming
previously grown AP-VOPO crystals under suitable pressure and
temperature conditions (3 GPa/800$^{\circ}$C) \cite{Gross02}.
Structure studies reveal \textit{phase purity} within the
experimental resolution of 2-3\% and give parameters in accordance
with published data \cite{Saito00}. ESR measurements yield a
$g$-factor of $g$ = 1.977 for $B \bot c$. The magnetic
susceptibility, measured by employing a Quantum Design SQUID
magnetometer, is very similar to the results in
ref.\,\cite{Johnston01}. A good description of our $\chi (T)$ data
for $B \bot c$ at 2-350\,K is obtained by using eq.\,(13a) in
ref.\,\cite{Johnston01}, the above $g$-factor, and the following
parameters: $J_1 / k_{B}$ = 127\,K, $J_2/ k_{B}$ = 111\,K, i.e.,
$\tilde{\alpha}$ = 0.873 and $\Delta / k_{B}$ = 31.8\,K, and a
concentration of 3.9$\%$ of paramagnetic $S$ = 1/2 moments per
mole V$^{4+}$ with an afm Weiss temperature $\Theta_{W}$ = (2
$\pm$ 0.6)\,K. These parameters, being in good agreement with
those given in ref.\,\cite{Johnston01}, also provide a consistent
description of the specific heat and thermal expansion data, see
below, and indicate that impurity effects are of no relevance in
the $T$ range discussed here.

The thermal expansion was measured at temperatures 1.5-200\,K
employing a high-resolution capacitive dilatometer \cite{Pott83}.
For the specific heat measurements at 2-30\,K, a home-built AC
calorimeter \cite{Mueller02} was used.

\begin{figure}
\begin{center}
  \includegraphics[width=\columnwidth]{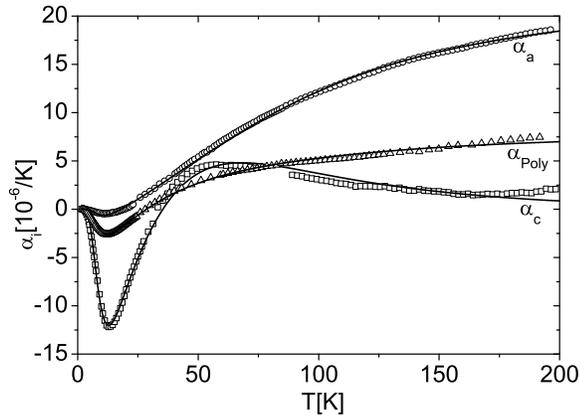}
  \caption{Expansion coefficients of single-crystalline HP-VOPO along the inter-chain $a$- (open circles),
  and intra-chain $c$-axis (open squares) and for polycrystalline material $\alpha_{poly}$ (open triangles).
  Solid lines are fits discussed in the text.}
  \label{Figalpha}
\end{center}
\end{figure}

Figure \ref{Figalpha} depicts the uniaxial expansivities
$\alpha_{i}(T) = l^{-1} dl/dT$ measured on an HP-VOPO single
crystal along ($i=c$) and perpendicular ($i=a$) to the spin-chain
direction, together with data for polycrystalline material
$\alpha_{poly}$. In accordance with the crystal structure, the
$\alpha_{i}$'s are strongly anisotropic. Particularly striking is
the anomalous in-chain $c$-axis expansivity, $\alpha_{c}$,
yielding a huge negative peak anomaly around 13\,K followed by a
change of sign and a maximum at 50\,K. At elevated temperatures $T
\gtrsim$ 150\,K, $\alpha_{c}$ becomes very small, indicating an
unusually small lattice contribution to $\alpha_{c}$. In contrast,
$\alpha_{a}$ and $\alpha_{poly}$ show a much larger lattice
expansivity with a small minimum at 13\,K and no obvious anomaly
at 50\,K.

\begin{figure}
\begin{center}
  \includegraphics[width=\columnwidth]{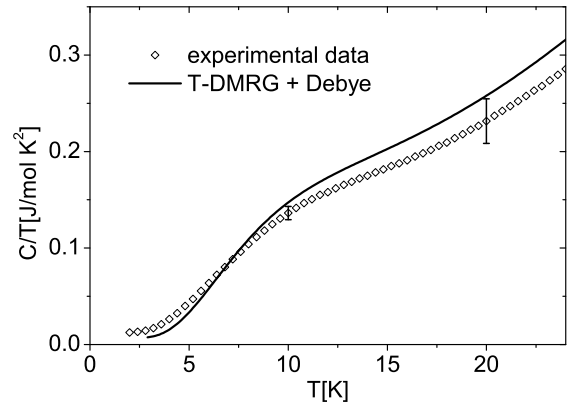}
  \caption{Specific heat as $C/T$ of single crystalline HP-VOPO ($\diamond$) of mass $m$ = 0.74\,mg.
  Solid line corresponds to model calculations described in the text.}
  \label{spec heat}
\end{center}
\end{figure}

As will be argued below, it is these two distinct features at
50\,K and 13\,K, dominating the intra-chain expansivity
$\alpha_{c}$, in which the two energy scales become manifested in
HP-VOPO. An indication for the low-energy scale, though much less
strongly pronounced, is also visible in the specific heat of
HP-VOPO, shown in Fig.\,\ref{spec heat}. The data reveal a weak
variation at low temperatures, consistent with a spin gap, and a
shoulder-like anomaly around 13\,K. The latter feature can be
discerned also in the $C(T)$ data on HP-VOPO by Azuma \textit{et
al.}\,\cite{Azuma99}, where it has been assigned to a
Schottky-type anomaly.

For the description of the magnetic contribution $C^{m}$ to the
specific heat of HP-VOPO in Fig.\,\ref{spec heat}, the density
matrix renormalization group approach for transfer matrices (T-DMRG)
has been used where the free energy of a one-dimensional quantum
system can be calculated by use of the Trotter-Suzuki mapping
\cite{Sirker02}. For all calculations we have retained between 64
and 100 states and used the Trotter parameter $\epsilon$ = 0.025 so
that the error of the Trotter-Suzuki mapping is of the order
$O(\epsilon^{2}) \approx 10^{-4}$, whereas the truncation error is
of the order $10^{-7}$. The phonon background $C^{ph}$ was
approximated by a Debye model with a Debye temperature $\Theta_{D}$
= 426\,K, which provides a reasonable description of the lattice
contribution to the thermal expansion, see below. Instead of fitting
the $C^{m}$ data with $J_{1}$ and $J_{2}$ as adjustable parameters,
requiring extensive computing time, T-DMRG calculations for $J_{1}/
k_{B}$ = 127\,K and $J_{2}/ k_{B}$ = 111\,K were used. As the solid
line in Fig.\,\ref{spec heat} demonstrates, this ansatz describes
the $C(T)$ data reasonably well. Most notably, it proves the hump
anomaly in $C(T)$ at low temperatures to be an intrinsic feature of
the weakly-dimerized afm S = 1/2 Heisenberg chain. The departure of
the solid line from the data at higher temperatures is assigned to
the inadequacy of the simple Debye model in accurately describing
the phonons in this temperature range.

\begin{figure}
  \includegraphics[width=\columnwidth]{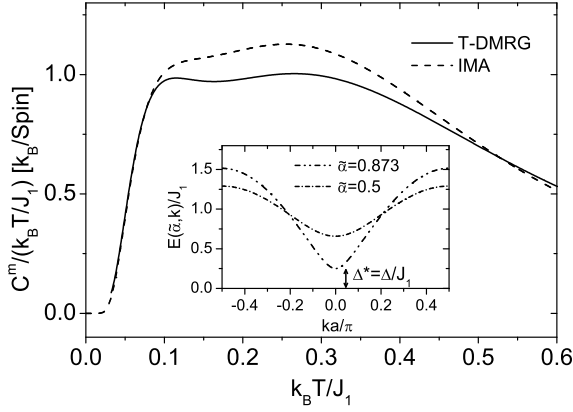}
  \caption{Magnetic specific heat divided by
  $k_{B} T/J_{1}$ vs. $k_{B}T / J_{1}$ for the afm $S$ = 1/2 alternating-exchange Heisenberg chain
  with $J_{1}/ k_{B}$ = 127\,K and $J_{2}/ k_{B}$ = 111\,K
  calculated by T-DMRG (solid line) and IMA (broken line). Inset shows
  one-magnon dispersion relations for the alternation parameter
  $\tilde{\alpha}$ = 0.873 and 0.5, calculated by using eq.\,(23a)
  in ref.\,\cite{Johnston00}.}
  \label{disprel}
\end{figure}

To gain more insight into the nature of the distinct
low-temperature anomalies in HP-VOPO, a simplified model, denoted
\textit{independent-magnon approximation} (IMA) in the following,
will be introduced, see also ref.\,\cite{Johnston00}. The
significance of the IMA is verified by comparison with the
accurate T-DMRG results. The thermodynamic calculations performed
within the IMA are based on a singlet ground state separated by an
energy gap $\Delta$ from a band of spin-1 excitations, cf.\ inset
to Fig.\,\ref{disprel}. To restrict the maximum number of excited
states to $2^{N}$, with $N$ the number of $S$ = 1/2 sites, the
occupancy of a one-magnon state with energy $E(k)$ and wave vector
$k$ is allowed to be either 0 or 1. In zero field, where the
one-magnon band is threefold degenerate, we obtain for the
magnetic contribution to the free energy per spin:

\begin{equation}\label{FE}
 F^{m} =
-\frac{k_BTd}{2\pi}\int_{-\pi/2d}^{\pi/2d}\ln\left(1+3\exp\left(-\frac{E(\tilde{\alpha},k)}{k_BT}\right)\right)\,dk.\
\end{equation}

Here $d$ is the average spin-spin distance along the chains. The
one-magnon dispersion $E(\tilde{\alpha}, k)$ was calculated to
high accuracy by Barnes \textit{et al.}\,\cite{Barnes99}. A fairly
simple parametrization, given by eq.\,(23a) in
ref.\,\cite{Johnston00}, is used here for numerical computations
of $F^{m}$, the magnetic entropy $S^{m} = -(\partial
F^{m}/\partial T)_{V}$ and $C^{m} = -T(\partial^{2}F^{m}/\partial
T^{2})_{V}$. Figure \ref{disprel} compares the so-derived $C^{m}$
with results from T-DMRG calculations. A representation $C/T$ is
chosen to make the low-temperature anomaly clearer. As expected,
the IMA nicely traces the T-DMRG results at low temperatures,
where the magnon density is small. In particular, it shows the
peak anomaly in $C^{m}/T$ at $k_{B}T/J_{1} \sim$ 0.1, a signature
of the low-energy scale. For temperatures $0.06 \leq k_{B}T/J_{1}
\leq 0.5$, however, the IMA tends to overestimate $C^{m}$ by about
10\% at maximum. The accompanied extra entropy is balanced at
higher temperatures, so that at $k_{B}T$ = 2$J_{1}$ (not shown),
both models reveal about 97 $\%$ of the full spin entropy.

\begin{figure}
\begin{center}
  \includegraphics[width=\columnwidth]{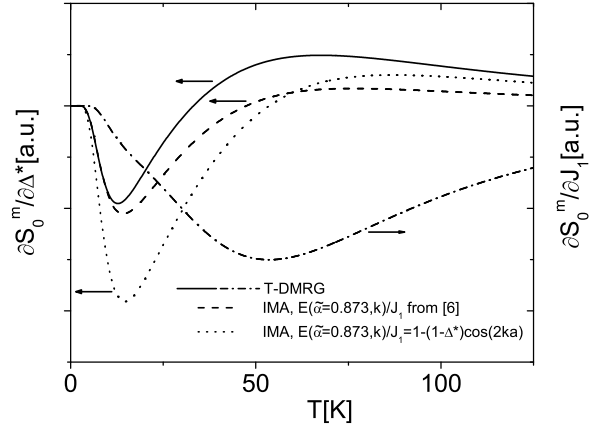}
  \caption{Temperature dependence of (i) $\partial S_{0}^{m} / \partial \Delta^{\ast}$
  (left scale) and (ii) $\partial S_{0}^{m}/ \partial J_{1}$ (right scale) for the
  alternating afm S = 1/2 Heisenberg chain with $J_{1}/k_{B}$ = 127\,K and $J_{2}/k_{B}$ = 111\,K,
  i.e.\, $\tilde{\alpha}$ = 0.873 and $\Delta/k_{B}$ = 31.8\,K. While
  for (i) T-DMRG results are compared with IMA results for different dispersion relations
  given in the figure, (ii) is calculated within T-DMRG.}
  \label{dSdepsilon}
\end{center}
\end{figure}

With these models at hand, a detailed description of the anomalous
thermal expansion of HP-VOPO is possible. If the entropy of a
system can be expressed as a sum of separate contributions $S =
\sum_j S^{j}$, the volume expansion coefficient can be written as
\cite{Barron80} $\beta = \sum_j\beta^j=\kappa_{T}\sum_j(\partial
S^{j}/ \partial V)_{T}$, with $\kappa_T = -V^{-1} (\partial V /
\partial P)_{T}$ the isothermal compressibility. This holds true
also for anisotropic systems as long as the volume changes occur
under hydrostatic conditions \cite{Barron80}. For HP-VOPO, the sum
$j$ runs over the magnetic ($j = m$) and phonon ($ph$)
contributions. The use of the Debye model and the \textit{scale
invariance} of $S^{m}$ imply that $S = S^{ph}(T/\Theta_{D}) +
S^{m}(T/J_{1}, \Delta^{\ast})$, where $\Delta^{\ast}$ = $\Delta
/J_{1}$. With $C^{ph}_{0} = -\Theta_{D}
\partial S^{ph}_{0}/\partial \Theta_{D}$ and $C^{m}_{0} =
-J_{1}\partial S^{m}_{0}/
\partial J_{1}$, the phonon and magnetic molar specific heat contributions, respectively, one finds

\begin{equation}\label{Beta}
\beta =
\frac{\kappa_{T}}{V_{0}}(-\Delta^{\ast}\Gamma^{\Delta^{\ast}}
\frac{\partial S^{m}_{0}}{\partial\Delta^{\ast}} + \Gamma^{J_{1}}
C^{m}_{0} + \Gamma^{ph} C^{ph}_{0}).
\end{equation}

Here $\Gamma^{\Delta^{\ast}} = -\partial ln \Delta^{\ast}/\partial
ln V$, $\Gamma^{J_{1}} = -\partial ln J_{1}/\partial ln V$ and
$\Gamma^{ph} = -\partial ln \Theta_{D} /\partial ln V$ define
different Gr\"{u}neisen parameters which measure the volume
dependence of the respective characteristic temperature or energy.
Note that as a consequence of $S^{m}$ being a function of the two
scaling variables $T/J_{1}$ and $\Delta^{\ast}$, $\beta$
(eq.\,\ref{Beta}) contains a contribution $\propto
\partial S^{m}_{0}/\partial\Delta^{\ast}$, which has no
counterpart in the magnetic specific heat, i.e., $\beta^{m}$ is no
longer proportional to $C^{m}$, viz. Gr\"{u}neisen-scaling does
not apply.

Equation \ref{Beta} can be generalized to the uniaxial
expansivities by $\alpha_i=\gamma_i^{\Delta^{\ast}} \partial
S^{m}_{0}/\partial\Delta^{\ast}+\gamma_i^{J_{1}} \partial
S^{m}_{0}/
\partial J_{1}+\gamma_i^{ph} \partial S^{ph}_{0}/ \partial
\Theta_{D}$, $i = a, b, c$. The prefactors
$\gamma_i^{\Delta^{\ast}}$, $\gamma_i^{J_{1}}$ and $\gamma_i^{ph}$
contain the respective strain dependencies of $\Delta^{\ast}$,
$J_{1}$ and $\Theta_{D}$, some combinations of elastic compliances
and the molar volume $V_{0}$, see \cite{Barron80}. Since these
quantities, along with the Gr\"{u}neisen parameters $\Gamma^{j}$
in eq.\,(\ref{Beta}), generally reveal only a very weak variation
with temperature, they can be treated as constants \cite{note2}.
In leading order, the $T$ dependence of $\alpha_{i}$ and $\beta$
is thus given by the derivatives of the entropy with respect to
the different energy scales.

The two magnetic terms of interest here, computed numerically
within the IMA (eq.\,(\ref{FE})) and T-DMRG for $J_{1}/ k_{B}$ =
127\,K and $J_{2}/ k_{B}$ = 111\,K, are displayed in
Fig.\,\ref{dSdepsilon} as a function of temperature. For $\partial
S^{m}_{0}/\partial J_{1}$ ($= -C^{m}_{0}/J_{1}$), a broad anomaly
is revealed around 50\,K, reflecting the short-range afm
correlations associated with $J_{1}$. In addition, $\partial
S^{m}_{0}/\partial J_{1}$ discloses a tiny feature around 13\,K,
corresponding to the low-temperature maximum in $C^{m}/T =
-J_{1}/T \cdot \partial S^{m}_{0}/\partial J_{1}$ in
Fig.\,\ref{disprel}, as a result of the gap formation. Since this
anomaly is much smaller and has the same sign as the one at 50\,K,
it cannot account for the change in sign and the large negative
peak anomaly in $\alpha_{c}$ at 13\,K. However, as shown in
Fig.\,\ref{dSdepsilon}, such an anomaly is revealed by $\partial
S^{m}_{0}/\partial \Delta^{\ast}$. This feature, constituting the
central result of this paper, reflects the sensitivity of the
magnetic entropy to changes in the spin gap $\Delta$, probed in a
most sensitive differential way by the expansivity. The effect can
be observed only in the weak-dimerization limit, where the energy
scales are clearly separated. Here the minimum in the one-magnon
dispersion is narrow so that the magnetic excitation spectrum is
very susceptible to small changes in $\Delta$, cf.\ inset to
Fig.\,\ref{disprel}. To simulate the effect of such changes on the
expansivity, comparative calculations of $\partial
S^{m}_{0}/\partial \Delta^{\ast}$ have been performed within the
IMA using different dispersion relations $E(\tilde{\alpha}, k)$
delimiting the spectrum to low energies, cf.\
Fig.\,\ref{dSdepsilon}. While for $E(\tilde{\alpha} = 0.873,k)$
shown in the inset to Fig.\,\ref{disprel}, the calculations within
the IMA (broken line in Fig.\,\ref{dSdepsilon}) nicely reproduce
the T-DMRG results (solid line) up to the minimum position, a
gapped $cos(2ka)$ dispersion with the same gap value gives rise to
an anomaly in $\partial S^{m}_{0}/\partial \Delta^{\ast}$ (dotted
line) which markedly deviates from the T-DMRG data. These results
highlight the extraordinarily high sensitivity of the expansivity
to peculiarities of the magnetic excitation spectrum associated
with the low energy scale.

Employing the prefactors $\gamma_{i}^{J_{1}}$,
$\gamma_{i}^{\Delta^{\ast}}$ and $\gamma_{i}^{ph}$ and
$\Theta_{D}$ as adjustable parameters, the expression for the
uniaxial expansivities can be used to fit the $\alpha_{a}$ and
$\alpha_{c}$ data in Fig.\,\ref{Figalpha} (solid lines). The good
quality of the fit to $\alpha_{a}$ for $T \gtrsim$ 70\,K, where
the phonon contribution dominates, validates the use of the Debye
model with $\Theta_{D}$ = 426\,K for satisfactorily describing the
phonons in HP-VOPO. The lattice contribution, however, plays a
minor role for $\alpha_{c}$. Here an excellent fit is obtained
essentially by the superposition of the two magnetic contributions
depicted in Fig.\,\ref{dSdepsilon}, properly scaled by
$\gamma_{c}^{J_{1}}$ and $\gamma_{c}^{\Delta^{\ast}}$. After
having assured the validity of our ansatz through these fits to
$\alpha_{c}$ and $\alpha_{a}$, a modeling of the volume
expansivity $\beta$ = $\alpha_{a}$ + $\alpha_{b}$ + $\alpha_{c}$
is possible by means of eq.\,(\ref{Beta}), cf.\ solid line through
$\alpha_{poly}$ = $\beta$/3 in Fig.\,\ref{Figalpha}. Employing a
rough estimate for the bulk modulus of $c_{B}$ = $\kappa_{T}^{-1}$
= (47 $\pm$ 6)\,GPa from $\Theta_{D}$ = 426\,K \cite{note3}, the
fit yields the following Gr\"{u}neisen parameters
$\Gamma^{\Delta^{\ast}}$ = -45 $\pm$ 6, $\Gamma^{J_{1}}$ = 1.3
$\pm$ 0.2 and $\Gamma^{ph}$ = 0.3 $\pm$ 0.1. Particularly striking
is the large $\Gamma^{\Delta}$ = $\Gamma^{\Delta^{\ast}}$ +
$\Gamma^{J_{1}}$ $\simeq$ -43 $\pm$ 6, reflecting a huge pressure
dependence of the spin gap $\partial \Delta / \partial P = \Delta
\cdot \kappa_{T} \cdot \Gamma^{\Delta}$ = -(28 $\pm$ 8)\,K/GPa. A
linear extrapolation to finite pressure implies that hydrostatic
pressure of about (1.1 $\pm$ 0.3)\,GPa is sufficient to close the
spin gap in HP-VOPO, i.e. to drive the system to the quantum
critical uniform limit.

In summary, the high-pressure variant of (VO)$_2$P$_2$O$_7$ has
been used to study the effects of two distinct energy scales of
the afm $S$ = 1/2 weakly-dimerized Heisenberg chain. We show that
$\partial S^{m}_{0}/\partial J_{1}$ accounts for the signatures in
the specific heat associated with short-range afm correlations and
spin-gap formation. However, this term cannot explain the large
low-temperature anomaly observed in the intra-chain expansivity.
The latter feature is well described by $\partial
S^{m}_{0}/\partial \Delta^{\ast}$, which probes the entropy
contours with respect to changes in the spin gap. This term,
inaccessible by specific heat, is observable only in the
weak-dimerization limit and reflects the high sensitivity of the
excitation spectrum, as regards small changes in $\Delta$. Our
results, which are based on the scale invariance of the entropy
$S^{m}(J_{1}, J_{2}, T) = S^{m}(T/J_{1}, J_{2}/J_{1})$, can be
generalized to cover the large family of magnetic multi-scale
systems described by Heisenberg-like models.

\begin{acknowledgments}
The authors acknowledge fruitful discussions with B. L\"{u}thi. The
T-DMRG calculations have been supported by the DFG under contracts
No.\ KL645/4-2 and GK1052.
\end{acknowledgments}


\begin{thebibliography}{Buravov 271}



\bibitem{Haldane83}F.D.M. Haldane, Phys. Rev. Lett. $\mathbf{50}$, 1153 (1983).
\bibitem{Affleck89}I. Affleck, J. Condens. Matter $\mathbf{1}$, 3047 (1989).
\bibitem{Dagotto96}E. Dagotto and T.M. Rice, Science $\mathbf{271}$, 618 (1996).
\bibitem{Oshikawa97}M. Oshikawa \textit{et al.}, Phys. Rev. Lett. $\mathbf{78}$, 1984 (1997).
\bibitem{Bonner79}J.C. Bonner \textit{et al.}, J. Appl. Phys. $\mathbf{50}$, 1810 (1979).
\bibitem{Johnston00}D.C. Johnston \textit{et al.}, Phys. Rev. B $\mathbf{61}$, 9558 (2000).


\bibitem{Signatures}An anomaly can also be
discerned in $\chi(T)$ calculated by various techniques, provided
$\tilde{\alpha} \gtrsim$ 0.8 \cite{Johnston00}.
\bibitem{Barron80}T.H. Barron \textit{et al.}, Adv. in Phys. $\mathbf{29}$, 609 (1980).

\bibitem{Azuma99}A. Azuma \textit{et al.}, Phys. Rev. B $\mathbf{60}$, 101454 (1999).
\bibitem{Johnston01}D.C. Johnston \textit{et al.}, Phys. Rev. B $\mathbf{64}$, 134403 (2001).
\bibitem{Garrett97a}A.W. Garrett \textit{et al.}, Phys. Rev. Lett. $\mathbf{79}$, 745 (1997).






\bibitem{Dasgupta00} O.K. Andersen and T. Saha-Dasgupta, Phys. Rev. B {\bf 62},
                     R16219 (2000), and references therein.




\bibitem{Gross02}C. Gross \textit{et al.}, High Pressure Research $\mathbf{22}$, 581 (2002).
\bibitem{Saito00} T. Saito \textit{et al.}, J. Solid State Chemistry {\bf 153}, 124 (2000).
\bibitem{Pott83}R. Pott \textit{et al.}, J. Phys. E $\mathbf{16}$, 444 (1983).
\bibitem{Mueller02}J. M\"{u}ller \textit{et al.}, Phys. Rev. B $\mathbf{65}$, R140509 (2002).
\bibitem{Sirker02}J. Sirker, A. Kl\"{u}mper, Europhys. Lett. {\bf
60}, 262 (2002).
\bibitem{Barnes03}T. Barnes, Phys. Rev. B $\mathbf{67}$, 024412 (2003).
\bibitem{Barnes99}T. Barnes \textit{et al.}, Phys. Rev. B $\mathbf{59}$, 11384 (1999).
\bibitem{note2}This is no longer valid close to the quantum critical point ($\Delta \rightarrow 0$), where the Gr\"{u}neisen parameter diverges \cite{Zhu03}.
\bibitem{Zhu03}L. Zhu \textit{et al.}, Phys. Rev. Lett. $\mathbf{91}$, 066404 (2003).
\bibitem{note3}This is possible within an isotropic approximation,
see G.A. Alers in \textit{Physical Acoustics}, Academic Press
(1965).


\end{thebibliography}

\end{document}